\documentstyle[preprint,prl,aps]{revtex}                              
\tightenlines
\begin{document}
\title{A mixed ultrasoft/normconserved pseudopotential scheme}
\author{K. Stokbro$^{1,2}$}
\address{$^{1}$Scuola Internazionale Superiore di Studi Avanzati,
 via Beirut 4, I-34014 Trieste, Italy. \\
$^{2}$Mikroelektronik Centret, Danmarks Tekniske Universitet, 
Bygning 345\o , DK-2800 Lyngby, Denmark.}

\maketitle

\begin{abstract}
A variant of the Vanderbilt ultrasoft pseudopotential scheme, where
the normconservation is released for only one or a few angular
channels, is presented. Within this scheme some difficulties of the
truly ultrasoft pseudopotentials are  overcome without
sacrificing the pseudopotential softness. i) Ghost states are easily 
avoided without including semicore shells. ii)  The ultrasoft
pseudo-charge-augmentation functions can be made more soft. iii) The
number of 
nonlocal operators is reduced. The scheme will be most useful for
transition metals, and the feasibility and accuracy of the scheme is
demonstrated for the 4d transition metal rhodium.
\end{abstract}
With the development of the Vanderbilt Ultrasoft-pseudopotential
(US)  technique\cite{Va90} it has become possible to describe
traditionally hard pseudopotential elements like transition metals
and first row elements with a modest plane-wave cutoff. However, for
some elements it is difficult to avoid the appearance of so-called
ghost states without including semicore states in the pseudopotential 
construction\cite{Va}. Furthermore, sometimes it is necessary
to describe the US pseudo-augmentation charge with a higher cutoff
than the 
pseudo-charge of the wave-function sum, and therefore so-called
double-grid  techniques have been 
developed\cite{LaPaCaLeVa93}. Both these aspects significantly reduce 
the computational efficiency of 
the US scheme, and it is desirable to find a simple scheme to 
avoid these difficulties.

In this report I will present a variant of the US
technique, where only the Normconservation(NC) is
released for some of the angular channels. The benefit from this
construction is that the semi-local potential of one of the chemical
active channels may be taken 
as the local potential. This reduces the number of projectors in the
US scheme and by choosing the semi-local
potential of the least bound reference state  as the local potential,
the  appearance of ghost states is effectively avoided. Furthermore,
since with this construction  there are less
pseudo-charge-augmentation functions, I have found the 
pseudo-charge smoothening procedure of Ref.~\cite{LaPaCaLeVa93} to
be more efficient. In the following I will briefly
review the main features
of the US scheme, and present the
modifications in a mixed
US/NC scheme. Finally, I will build a US/NC pseudopotential for Rh
and compare the
transferability and computational efficiency 
 with that of a NC pseudopotential and an US pseudopotential with
semicore shells.

In the US scheme the pseudo-wave-functions do not obey a NC condition
and it is this feature which makes it
possible to construct pseudopotentials with a modest plane-wave cutoff
($\leq 30$Ry) for traditionally hard pseudopotential elements. 
 To make the pseudopotential retain the first order scattering
properties of the all-electron potential, as is the case of the NC
pseudopotentials, the pseudo-wave-functions have to obey a
generalized eigenvalue equation, where the missing norm enters in the
overlap matrix. Furthermore, the density cannot simply be constructed
from the pseudo-wave-functions, an augmentation charge has to be added
in order to retain the correct electro-static potential in the
interstitial region. Given a set of occupied pseudo-wave-functions
$\left\{\Phi_{\alpha}\right\}$, the augmentation charge is given by
\begin{eqnarray}
\rho^{aug} ({\bf r}) & = & \sum_{\alpha,n,m} 
 \langle\Phi_{\alpha}|\chi_n\rangle Q_{nm}({\bf r})
\langle\chi_m|\Phi_{\alpha}\rangle \; , \\
\label{eq:aug}
Q_{nm}({\bf r}) & = & \psi^*_n({\bf r}) \psi_m({\bf r}) -
\phi^*_n({\bf r}) \phi_m({\bf r})  \;, 
\end{eqnarray}
where $\psi_n$ are the reference all-electron atomic wave-functions, $\phi_n$
the corresponding reference atomic  pseudo-wave-functions, and $\chi_m$ are
projectors upon the reference atomic pseudo-wave-functions. 

With this choice for the augmentation charge the total pseudo density
does not only
have the same norm as the all-electron density, as in the 
case of NC pseudopotentials, but is identical to the all-electron
density in the 
case where the  set of reference states inside each core form a
complete basis for the 
all-electron wave-functions and the pseudo-wave-functions,
 i.e. $\Psi_{\alpha} = \sum_na_n^{\alpha} \psi_n$ and $\Phi_\alpha =
\sum_n b_n^{\alpha} \phi_n$. To see this observe that 
since the pseudo and all-electron wave-functions and their
radial derivatives coincide at the core radius they must have the same
expansion in the reference states, thus $a_n^{\alpha}= b_n^{\alpha} $
which implies that
$\sum_{\alpha}|\Psi_{\alpha}|^2=\sum_{\alpha}|\Phi_{\alpha}|^2+\rho^{aug}$.

Therefore, in the US scheme
not only the electro-static mono-pole of the core region is correct,
but also higher order poles are described rather accurately.
Another feature of the US scheme is the possibility to include
several reference states for each angular channel. 

In the NC scheme it is common practise to take one of the semi-local
potentials as the local potential and thereby reduce the number of
semi-local channels that have to be described. Furthermore, by
choosing the semi-local potential of the least
bound reference state one can effectively avoid the appearance of
ghost states\cite{Bl90,GoStSc91} in a Kleinmann-Bylander\cite{KlBy82}
implementation. A similar choice for the local potential in the
ultrasoft scheme would produce a very poor pseudopotential, since
due to the missing norm of the corresponding pseudo-wave-function the
local potential would only have the zero-order scattering properties
of the all-electron potential. Instead the common practise is to
construct a semi-local potential for the first angular channel which is
not chemical active and use that for the local potential, for
instance, for the 
transition metals 
the semi-local potential of the $f$-channel is usually used for the
local 
potential. However, for most transition metals this will produce a
local 
potential with bound states far below the $s$- and $p$-reference
eigenstates, and the appearance of ghost states is therefore
unavoidable\cite{GoStSc91}. A solution to this problem 
is to include $s$- and $p$-semicore shells in the pseudopotential
construction, since these new
reference eigenstates usually are below the bound states of the local
potential the ghost-states will disappear. The inclusion of semicore
shells has the benefit that  
the accuracy of the pseudopotential gets comparable to
all-electron calculations\cite{KiVa94}, however, it
also increases the computer time and memory requirements of the
calculation substantially. 

As an alternative I propose a mixed US/NC scheme, where the 
normconservation is only  released for the hard channels,  and the
semi-local 
potential of one of the soft channels may thereby be taken as the
local potential. For most  transition metals the normconservation will
 only have to be released for the $d$-channel, and the semi-local
potential of the  $p$-channel can therefore be taken as the local
potential, thereby avoiding the appearance of ghost-states. The
pseudo-charge-augmentation functions $Q_{nm}$ of Eq.~(\ref{eq:aug}) I
now define as  
\begin{equation}
Q_{nm}({\bf r})  =  \xi^*_n({\bf r}) \xi_m({\bf r}) - \phi^*_n({\bf
r}) \phi_m({\bf r}),
\end{equation}
where $\xi=\psi$ for the US channels and $\xi=\phi$ for the NC
channels, and only  contributions from  nonlocal channels are
included, i.e.  for most transition metals  only the $Q_{dd}$ and
$Q_{sd}$ elements will contribute to the augmentation charge.

In the following I will construct a scalar-relativistic
 mixed US/NC pseudopotential for the 4d transition metal Rh,  show that the
error introduced by  the choice for the augmentation charge is minute,
and compare the transferability and
computational efficiency with a  NC 
pseudopotential and a US pseudopotential with semicore states.

To construct the NC pseudopotential (denoted NC(9)) I used the
procedure suggested by 
Troullier-Martins\cite{TrMa91}, with core radii 2.53, 2.53 and
1.39 (a.u.) for the $s-$, $p$- and $d$-channel, respectively. 
As suggested in Ref.~\cite{BaHaSc82}, only the semi-local
$s-$ and $d$-potentials were constructed in the
atomic ground-state configuration $4d^85s^1$, while
the semi-local $p$-potential was  constructed in the $4d^75s^{0.75}p^{0.25}$
configuration.  Figure~1  shows the pseudo wave-functions and
their  Fourier transforms. From the latter it can be seen that
 the $s$- and $p$-wave-functions
are converged at $\approx 20-25$ Ry, while the $d$-wave-function must be
described with a cutoff of $\approx 50-60$ Ry. From this
pseudopotential  the mixed US/NC pseudopotential (denoted
US/NC(9)) was constructed  by 
releasing the norm of the $d$-wave-function, increasing its core
radius to 1.6 a.u, and then   using the smoothening procedure 
of Ref.~\cite{RaRaKaJo90,KiVa94} to generate  $d$-ultra-soft
pseudo-wave-functions
for two reference states, where
one has the atomic eigenvalue (-0.4518 Ry) and the other
the energy (0.2 Ry). With this construction 
 the $d$-pseudo-wave-function can  be described
with a cutoff  of $\approx 25-30$ Ry,  as seen from its Fourier
transform shown in Fig.~1b(dotted line). Finally, I have constructed an
US pseudopotential including  $4s$ and $4p$ semicore shells (denoted
US(17)), using  2 reference states for each channel, and  a
core-radius of 2.0, 2.0 and 1.6 for the $s$-, $p$-, and $d$-channel, respectively.
The US(17) pseudopotential is also converged at $\approx 25-30$ Ry.

To generate pseudo-charge-augmentation functions I have used the
procedure described in Ref.~\cite{LaPaCaLeVa93,KiVa94}, where
the augmentation functions are replaced by L-dependent counterparts
$Q^L_{nm}$,
\begin{equation}
Q_{nm}({\bf r}) = \sum_{LM} c_{LM}^{nm}Y_{LM}(\hat{r})Q^L_{nm}(r),
\end{equation}
which are smoothened inside a core radius, $r_{in}^L$, subject to the
condition that the Lth moment of the electron charge density is
conserved. One difficulty with this method is that if the core radius
$r_{in}^0$ is extended beyond a certain radius the
smoothening procedure starts to develop negative
sections in the $Q_{nm}^{0}(r)$ terms, leading to negative
pseudo-charge densities. I have found that this difficulty
is related to the different node structure of the  radial all-electron
wave functions, and that a  larger $r_{in}^0$ core radius can be chosen when there
is only one angular channel.  Figure~2  shows the  $Q_{ss}^0$, $Q_{pp}^0$
and $Q_{dd}^0$ pseudo-charge-augmentation functions of the US(17) pseudopotential
and the  $Q_{dd}^{0}$ pseudo-charge-augmentation function of the US/NC(9)
pseudopotential. For the US(17) pseudopotential the core radius 
could not be extended beyond $r_{in}^0=0.6$ a.u., while the US/NC(9)
core radius was $r_{in}^0=0.9$ a.u. giving  softer
pseudo-charge-augmentation functions.

Table~1 shows the transferability of the US/NC(9), NC(9) and
US(17) pseudopotentials  for atomic, bulk and surface
properties of Rh. First notice  the very high quality of the
US(17) pseudopotential, having an accuracy essentially identical 
to all-electron calculations. Notice also the discrepancy (due to the
local-density approximation) between
the all-electron calculation and the experimental results for the bulk
properties of Rh, which warns that the quality of a pseudopotential
should never be judged by comparison with experimental data. 
The atomic calculations show that
the main difference between the NC(9) and the US/NC(9)
pseudopotentials, is the latter's enhanced description of the
$d$-electrons, which is due to the use of two reference states. 

Table~1c shows  the CPU time used to obtain the
self-consistent charge density of the non-relaxed surface for the three
different pseudopotential schemes. The comparison  reveals
that the US/NC(9) pseudopotential is 3.8 times faster that the US(17)
pseudopotential and 3.4 times faster than the NC(9) pseudopotential.
For  large systems the orthogonalization step is the most time-consuming
part of a plane-wave program, and this scales as $O(NM^2)$, where $N$ is
the number of plane-waves and $M$ the number of occupied bands. In the
present case this gives a time factor relative to the US/NC(9)
pseudopotential of $(17/9)^2\approx 3.6$ for the
US(17) pseudopotential and  $2^{\frac{3}{2}}\approx2.8$ for
the NC(9) pseudopotential. Another important issue is the memory
requirement which scales as $O(NM)$, which results in a factor
$17/9\approx1.9$ for the  US(9) pseudopotential and a factor
$2^{\frac{3}{2}}\approx2.8$ for
the NC(9) pseudopotential relative to the US/NC(9) pseudopotential.

To test whatever the neglect of the $Q_{pd}$ pseudo-charge-augmentation functions in the
US/NC scheme introduce any significant error, I have
constructed a US(17)$^*$ pseudopotential with $Q_{sp}=Q_{pd}=Q_{sd}=0$.
These terms are most important for asymmetric systems, however, the
calculated Rh(110) surface properties are almost unaffected by this
approximation(see Table~1c). Furthermore, I have found that different
variations of the US/NC scheme as including two $s$-reference states,
using the atomic ground state for the $p$-reference configuration or
including the $Q_{ss}$ term, do not change  the quality of the
pseudopotential significantly. The discrepancy between the US(17) and
US/NC(9) pseudopotentials is therefore mainly due to the latter's
neglect of semicore states.

In conclusion I have presented a simple scheme for
constructing  a  mixed normconserved/ultrasoft pseudopotential
using as a starting point a normconserved pseudopotential. The
resulting pseudopotential scheme is more
 accurate and computationally efficient than the initial normconserved
pseudopotential. Compared to truly ultrasoft schemes, ghost states
are avoided without including semicore shells and the ultrasoft
pseudo-augmentation charge is more easily described with the same
cutoff as the wave-function pseudo-charge. 

\acknowledgements
I acknowledge D. Vanderbilt for valuable discussion and
for providing an atomic program for
the US pseudopotential construction, and P. Giannozzi for help
with constructing the NC pseudopotential. I also thanks V. Fiorentini for
the  all-electron data for the bulk properties of Rh, and 
S. Baroni for
directing me towards the ultrasoft method and for valuable discussions
concerning the method. This work was done under support of the Danish
research councils, EEC contract ERBCHBGCT 920180,
EEC contract ERBCHRXCT 930342, and  CNR project   Supaltemp.

\begin{figure}
\caption{a) The real-space all-electron wave-functions of Rh
and the pseudo-wave-functions of
the NC(9) and US/NC(9) pseudopotentials. 
b) The Fourier-transformed pseudo-wave-functions.}
\end{figure}

\begin{figure}
\caption{The Fourier-transformed pseudo-charge-augmentation functions of the
US(17) and the US/NC(9) pseudopotential.}
\end{figure}

\mediumtext
\begin{table}
\caption{Comparison of the three rhodium pseudopotentials.  US(17) is 
the ultrasoft pseudopotential including semicore shells, NC(9) is the
normconserved pseudopotential and  US/NC(9) is the mixed
normconserved/ultrasoft scheme. In all the calculations, including the
atomic calculations of part a), the pseudopotentials are used in a
non-local separable form. a) The calculated
atomic-eigenvalues in different reference configurations
and their deviation from the all-electron values. b) The  lattice
constant ($a_{0}$), bulk modulus (B), derivative of the bulk modulus
(B'), and cohesive energy of rhodium. c) The surface energy
($E_{surf}$), workfunction($W$), and interlayer relaxation
($\Delta_{12},\Delta_{23}$ ) of the Rh(110) surface. The US(17)$^*$
pseudopotential neglects pseudo-charge-augmentation functions which couple
different angular channels, i.e. $Q_{sp}=Q_{pd}=Q_{sd}=0$. }

\begin{tabular}{lccccccc}
\multicolumn{8}{l}{a) Atomic properties of Rh}   \\
\multicolumn{1}{l}{pseudo }  & 
\multicolumn{1}{c}{configuration }  & 
\multicolumn{1}{c}{$4d$ [Ry]}  & 
\multicolumn{1}{c}{$\Delta(4d)$ [mRy]}  & 
\multicolumn{1}{c}{$5s$ [Ry]}  & 
\multicolumn{1}{c}{$\Delta(5s)$ [mRy]}  & 
\multicolumn{1}{c}{$5p$ [Ry]}  & 
\multicolumn{1}{c}{$\Delta(5p)$ [mRy]}  \\
\tableline
US(17) & $4d^95s^05p^0$ &  -0.28267& 0.0 & -0.26864 & 0.0 & -0.03501 & 0.0 \\
US/NC(9) & $4d^95s^05p^0$  & -0.28155 & 1.2 & -0.27227 & -3.6 & -0.03798 &
-3.0 \\ 
NC(9) & $4d^95s^05p^0$  & -0.27479 & 7.9 & -0.27022 & -1.6 & -0.03883 & -3.8 \\ 
\tableline
US(17) & $4d^75s^15p^0$ & -1.24280 &  0.1 & -0.91783 & 0.0 & -0.53999 &
0.1 \\
US/NC(9)  & $4d^75s^15p^0$ & -1.24384 & -0.9 & -0.91472 & 3.1 & -0.53760 &
2.5 \\
NC(9)  & $4d^75s^15p^0$ & -1.25787 & -15.0 & -0.91985 & -2.0 & -0.53800 &
2.1 \\
\end{tabular}
\begin{tabular}{lccccc}
\multicolumn{6}{l}{b) Bulk properties of Rh}\\
\multicolumn{1}{l}{pseudo }  & 
\multicolumn{1}{c}{cutoff [Ry] }  & 
\multicolumn{1}{c}{$a_0$ [\AA] }  & 
\multicolumn{1}{c}{B [Mbar]}  & 
\multicolumn{1}{c}{B' }  & 
\multicolumn{1}{c}{E$_{coh}$[Ry/atom]}  \\
\tableline
Exp. & &  3.80$^a$ & 2.76$^a$ &  & 0.42$^a$ \\
All El. & & 3.74$^b$ & 3.46$^b$ & 7.3$^b$ & 0.569$^b$ \\
US(17) & 30 &  3.75 & 3.44 & 7.8 & 0.636 \\
US/NC(9) & 30 &  3.81 & 3.16 & 5.6 & 0.609 \\
NC(9) & 60 &  3.86 & 2.98 & 5.5 & 0.594 \\
\end{tabular}
\begin{tabular}{lcccccc}
\multicolumn{6}{l}{c) Surface properties of Rh(110)}\\
\multicolumn{1}{l}{pseudo }  & 
\multicolumn{1}{c}{Cutoff [Ry] }  & 
\multicolumn{1}{c}{CPU time [s] }  & 
\multicolumn{1}{c}{E$_{surf}$ [eV/atom] }  & 
\multicolumn{1}{c}{W [eV] }  & 
\multicolumn{1}{c}{$\Delta d_{12}$ [\%]}  & 
\multicolumn{1}{c}{$\Delta d_{23}$ [\%]}  \\
\tableline
Exp. & & & & 4.98$^c$& -6.9$^d$ & 1.9$^d$ \\
US(17) & 30 & 5796 &  1.73 & 4.77 & -12.1 & 4.9 \\
US/NC(9) & 30 &  1514 & 1.92 & 4.99 & -10.2 & 2.5 \\
NC(9) & 60 & 5162 & 1.87 & 4.98 & -10.5 & 2.6 \\
US(17)$^*$& 30 & 5796 &  1.73 & 4.78 & -12.1 & 4.9 \\
\end{tabular}
$^a$Reference~\protect\cite{We95}
$^b$Reference~\protect\cite{fiore}
$^c$Reference~\protect\cite{NiBoSa74}(polycrystaline)
$^d$Reference~\protect\cite{NiBiHaHeMu87}
\end{table}


\begin{thebibliography}{10}

\bibitem{Va90}
D. Vanderbilt, Phys. Rev. B {\bf 41},  7892  (1990).

\bibitem{Va}
D. Vanderbilt, {\it Priv. Comm.}.


\bibitem{LaPaCaLeVa93}
K. Laasonen {\it et~al.}, Phys. Rev. B {\bf 47},  10142  (1993).

\bibitem{Bl90}
P.~E. Bl\"{o}chl, Phys. Rev. B {\bf 41},  5414  (1990).

\bibitem{GoStSc91}
X. Gonze, R. Stumpf, and M. Scheffler, Phys. Rev. B {\bf 44},  8503  (1991).

\bibitem{KlBy82}
L. Kleinman and D.~M. Bylander, Phys. Rev. Lett. {\bf 48},  1425  (1982).

\bibitem{KiVa94}
R.~D. King-Smith and D. Vanderbilt, Phys. Rev. B {\bf 49},  5828  (1994).

\bibitem{TrMa91}
N. Troullier and J.~L. Martins, Phys. Rev. B {\bf 43},  1993  (1991).

\bibitem{BaHaSc82}
G.~B. Bachelet, D.~R. Hamann, and M. Schl\"{u}ter, Phys. Rev. B {\bf 26},  4199
   (1982).

\bibitem{RaRaKaJo90}
A.~M. Rappe, K.~M. Rabe, E. Kaxiras, and J.~D. Joannopoulos, Phys. Rev. B {\bf
  41},  1227  (1990).

\bibitem{We95}
{\it CRC Handbook of Chemistry and Physics}, 75$^{th}$ Edition, Editor
D. R. Lide, (CRC Press, 1994).

\bibitem{fiore}
V. Fiorentini, {\it Priv. Comm.}. The calculations were performed
with the full-potential Linear-Muffin-Tin-Orbital(LMTO) method,
treating the $4s$, $4p$, $4d$, $5s$ and $5p$ states  as valence states, and relaxing the core states.

\bibitem{Gs64}
K.~A. Gschneider, Solid State Phys. {\bf 16},  276  (1964).

\bibitem{NiBiHaHeMu87}
W. Nichtl {\it et~al.}, Surf. Sci. {\bf 188},  L729  (1987).

\bibitem{NiBoSa74}
Nieuwenhuys, Bouwman, and Sachtler, Thin Solid Films {\bf 21},  51  (1974).

\end{thebibliography}
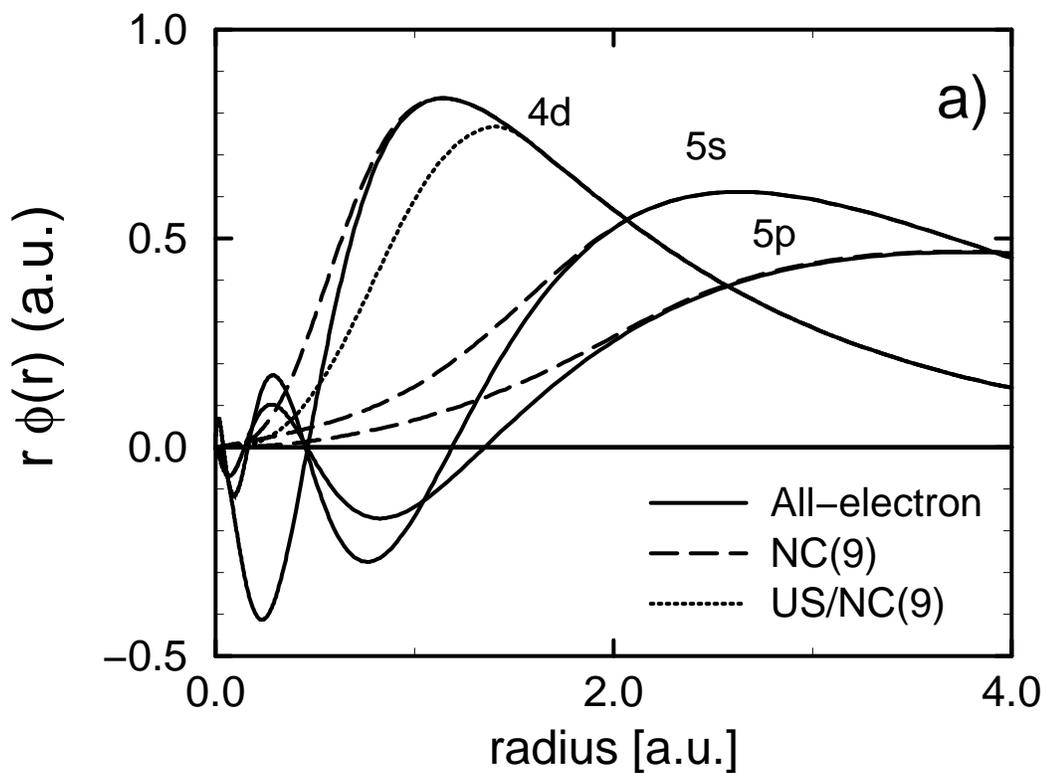
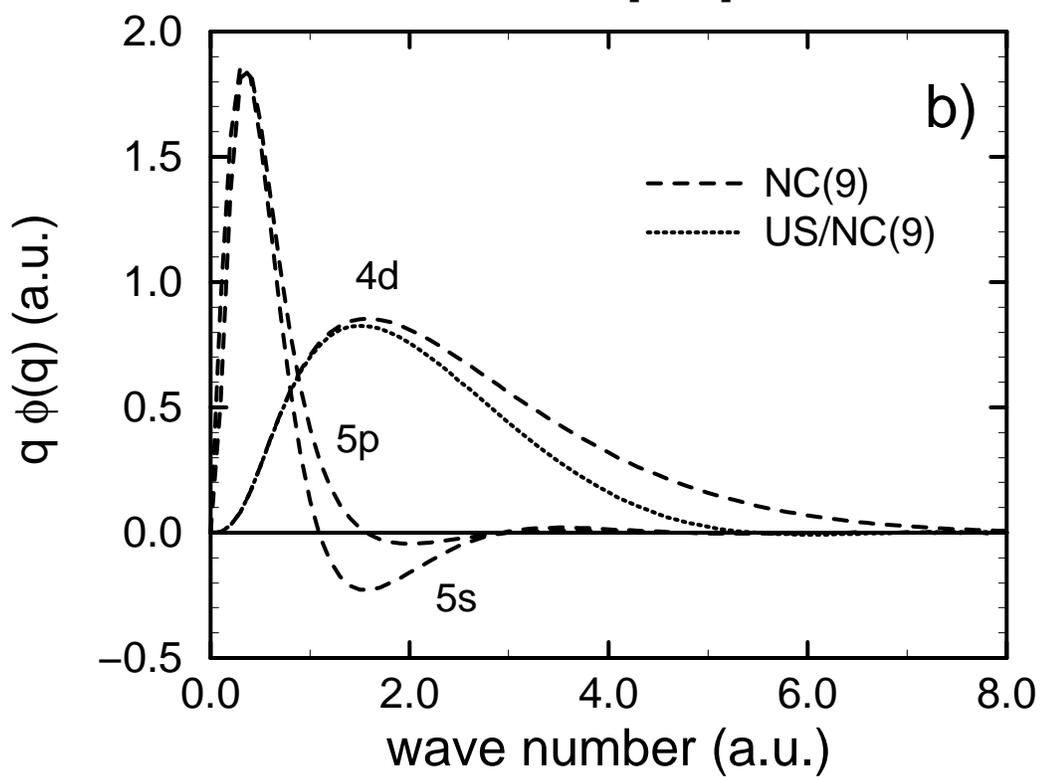
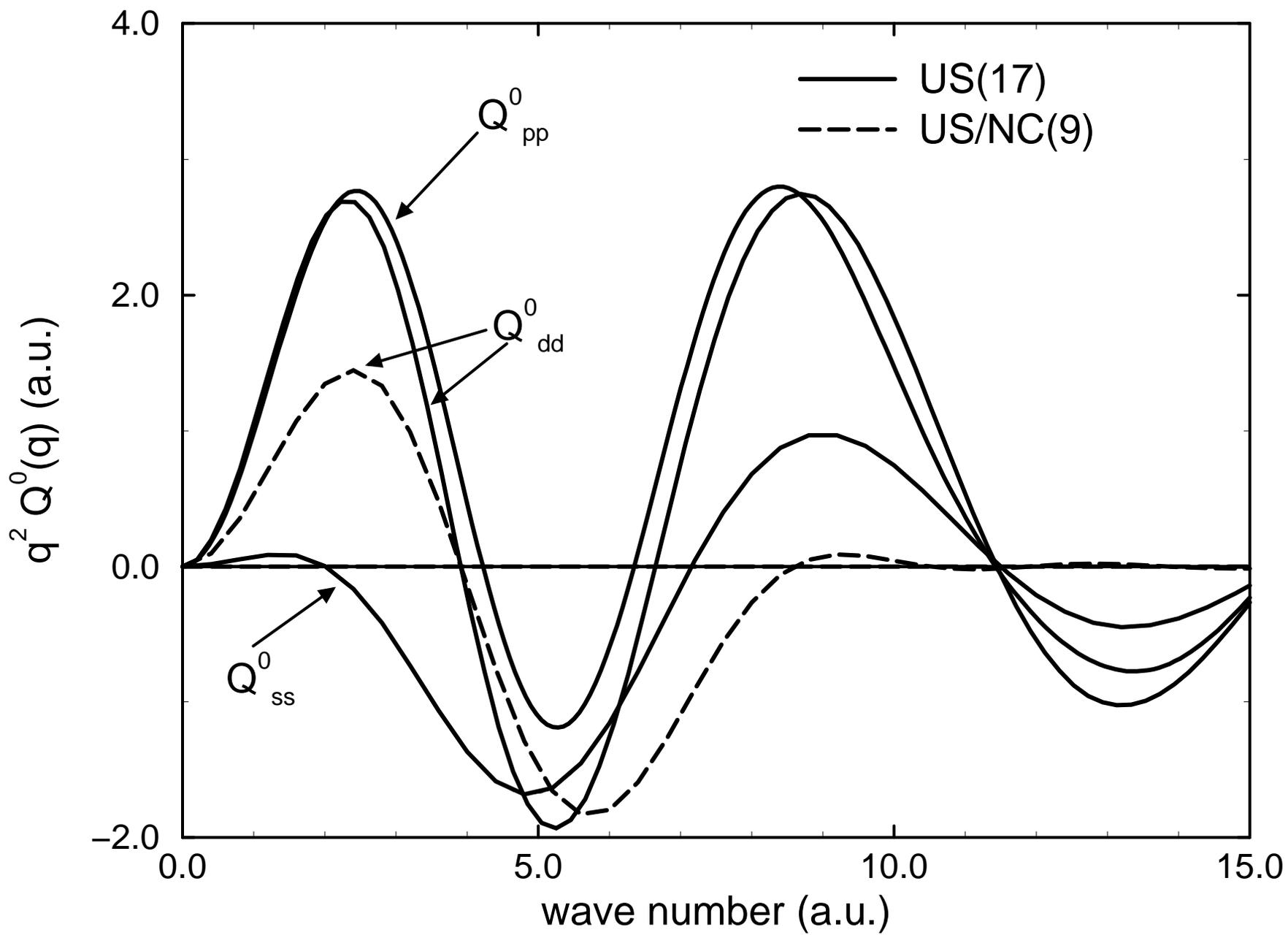
\end{document}